# Inter-dimensional optical isospectrality inspired by graph networks

Sunkyu Yu,[1] Xianji Piao,[1] Jiho Hong,[1] and Namkyoo Park[1,*]

[1]*Photonic Systems Laboratory, Department of Electrical and Computer Engineering, Seoul National University, Seoul 08826, Korea*
*Corresponding author: nkpark@snu.ac.kr*

**A network picture has been applied to various physical and biological systems to understand their governing mechanisms intuitively. Utilizing discretization schemes, both electrical and optical materials can also be interpreted as abstract 'graph' networks composed of couplings (edges) between local elements (vertices), which define the correlation between material structures and wave flows. Nonetheless, the fertile structural degrees of freedom in graph theory have not been fully exploited in physics owing to the suppressed long-range interaction between far-off elements. Here, by exploiting the mathematical similarity between Hamiltonians in different dimensions, we propose the design of reduced-dimensional optical structures that perfectly preserve the level statistics of disordered graph networks with significant long-range coupling. We show that the disorder-induced removal of the level degeneracy in high-degree networks allows their isospectral projection to one-dimensional structures without any disconnection. This inter-dimensional isospectrality between high- and low-degree graph-like structures enables the ultimate simplification of broadband multilevel devices, from three- to one-dimensional structures.**

Random scattering provides the spread of momentum and energy components of waves [1-3], differentiating disordered materials from periodic or quasiperiodic materials [4,5]. Due to the critical needs in light harvesting [6], robust bandgaps [7,8], and ultrafast optics [9], disordered optical materials have also been intensively studied to exploit their broadband responses. Although one- (1D) [8,10], two- (2D) [1,11,12], and three-dimensional (3D) [13] disordered structures have been considered promising candidates for broadband and omnidirectional operations, the deliberate control of disordered structures [1,3,7,8,14] has been achieved only in 1D or 2D structures, owing to the difficulty of manipulating 3D randomness.

To understand the role of the 'dimension' in optics, we can employ the interdisciplinary viewpoint from network theory. For the structures composed of coupled resonances, light flows inside those can be interpreted as signal transport over graph networks [14-17]. Reciprocity constructs the undirected network [14,18], whose vertices and edges denote resonances and coupling, respectively. The strength of disorder then corresponds to the graph irregularity, and the dimension of material structures determines the 'degree' of graphs, which quantifies the number of coupling paths in space. This graph-based perspective reveals that optical structures in 3D real space cover only restricted parts of general graph theory [18], originating from the difficulty in connecting far-off vertices in real space. At most, nearest-neighbor and next-nearest-neighbor coupling [19] have been considered in the light transport.

Here, we focus on the real-space reproduction of the level statistics in hypothetical optical networks of arbitrary couplings, consequently deriving the 'isospectrality' between the structures of different dimensions. We demonstrate that as similar to supersymmetric techniques for isospectral potentials [8,20-24], the Hamiltonian similarity allows the isospectral reduction of network degrees, forming the reduced dimensionality in real-space designs. The condition of a fully connected structure with a lower dimension and preserved level statistics is then investigated, showing that a 2D/3D disordered structure even with all orders of couplings can always be transformed into a single 1D structure with nearest-neighbor coupling. Our approach enables the 1D replication of the spectral information of 3D disordered structures, allowing the simplification of broadband multilevel devices, such as multimode waveguides [21], resonators [25], and bandgaps [8], and the inter-dimensional interaction from the global phase matching condition.

Consider the example of optical networks (Fig. 1a) and its graph representation (Fig. 1b). The network is obtained by the Watts-Strogatz rewiring [26] of a regular graph; each edge in the graph network which is selected in turn are reconnected with the rewiring probability $p$, to a vertex chosen uniformly at random (number of vertices $N$ = 21, graph degree $k$ = 4, and rewiring probability $p$ = 0.5). Each vertex represents an optical local mode [10,15,16], whereas the edge between vertices denotes a 'hypothetical' coupling between elements, including significant long-range coupling. As an illustrative case, we begin with the non-weighted network [18], whose edges (or vertices) have equal coupling strength (or resonance). We note that the proposed $N$-body system is hypothetical because the long-range coupling between far-off elements is suppressed for local modes.

The level statistics of the proposed $N$-body graph is determined by the eigenvalue equation $H\psi = E\psi$ [10,15] of

$$\frac{d}{dt}\psi_v = i\rho_o \psi_v + \sum_{w \neq v} i\kappa(v,w) \cdot \psi_w,$$

where $v = 1,2,\ldots,N$ is the vertex number, $\psi_v$ is the electromagnetic field at the $v^{th}$ element, $\rho_o$ is the self-energy of each element, and $\kappa(v,w)$ is the coupling coefficient between the $v^{th}$ and $w^{th}$ elements for the randomly-rewired graph ($\kappa(v,w) = \kappa_o$ for connected $v$-$w$ and $\kappa(v,w) = 0$ for disconnected $v$-$w$ vertices). The system Hamiltonian $H$ is symmetric ($H = H^T$) due to the reciprocity $\kappa(v,w) = \kappa(w,v)$.

For the Watts-Strogatz-rewired [26] high-degree graph, the off-diagonal part of its Hamiltonian $H$ is randomly filled with $\kappa_o$ or zero, in contrast to the sparse and quasi-diagonal Hamiltonians of real-space structures. To transform this hypothetical $H$ into a realistic form, we focus on the derivation of the 'effective' Hamiltonian, which includes only nearest-neighbor couplings for a 1D structure but possesses identical level statistics. We then apply the similarity between symmetric Hamiltonians based on Householder transformation [27]. For the orthogonal matrix $M^TM = I$, the eigenvalue equation can be cast in the form of $MH(M^TM)\psi = M(E\psi)$, deriving the isospectral transformation as $H_i(M\psi) = (MHM^T) \cdot (M\psi) = E(M\psi)$, where $H_i = MHM^T$ is the transformed Hamiltonian with an identical eigenspectrum $E$ and transformed eigenstates $M\psi$. By applying the tridiagonalization technique [27] from Householder transformation, the sequential transformation by orthogonal matrices as $H_{iq} = (M_q M_{q-1} \cdots$

$M_2M_1) \cdot H \cdot (M_qM_{q-1} \cdots M_2M_1)^T$ can remove off-diagonal terms of $H$ in order, except its tridiagonal components, when $M_q$ has the form of

$$M_q = \begin{pmatrix} I_q & O_q \\ O_q & L_{N-q} \end{pmatrix},$$

where $I_q$ and $O_q$ denote the $q \times q$ identity and the null matrix, respectively, $L_{N-q}$ has the Householder form [27] of $L_{N-q} = I_{N-q} - 2uu^T$, and $u$ is the unit normal vector which defines the reflection plane for the Householder transformation. Figure 1c shows the serial reduction of graph edges from the sequential transformation. The transformation with $M = M_{N-2}M_{N-3} \cdots M_2M_1$ finally derives the effectively 'weighted' graph with the degree $k = 2$ (Fig. 1c), corresponding to a finite 1D structure in real space with the tridiagonalized Hamiltonian $H_i = MHM^T$. We thus prove that a graph network, even one that includes significant long-range coupling (Fig. 1b), has a 1D isospectral partner structure which includes only nearest-neighbor couplings (Fig. 1d from the final graph of Fig. 1c; see Fig. 1e for its self-energy and coupling strength distribution). However, achieving this isospectral 1D realization of a network does not guarantee that the spectrum of the network can be reproduced fully by a 'single' 1D structure, as shown in the disconnection in Fig. 1d (black arrow).

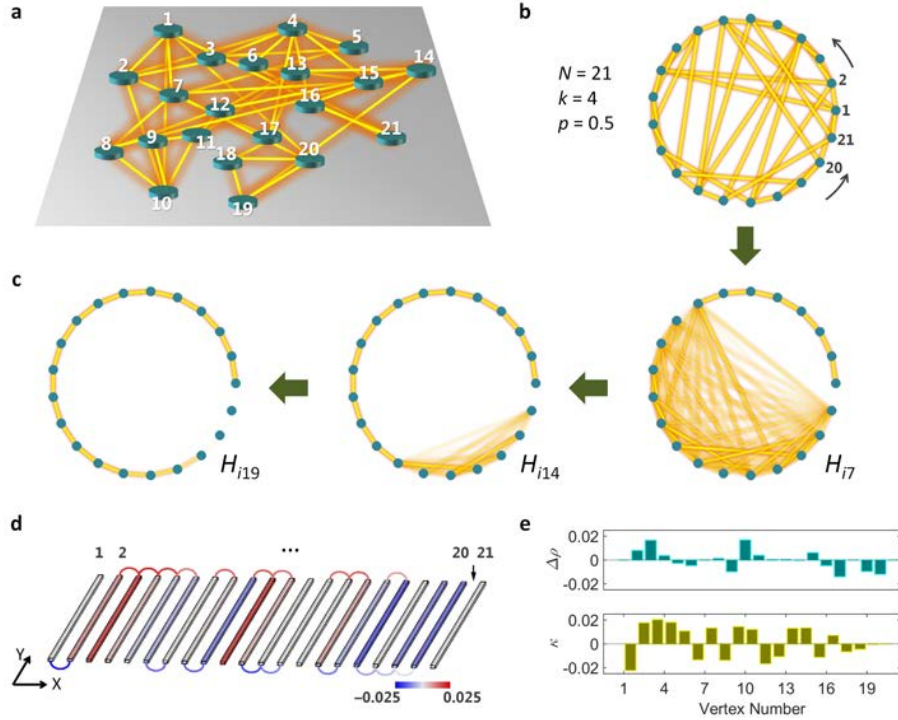

Fig. 1. Network-inspired design of 1D structure with a 'similar' Hamiltonian. (a) A schematic of an optical network ($N = 21$, $k = 4$, and $p = 0.5$), composed of hypothetically coupled resonances. (b) The graph representation of the structure. (c) The variation of network connections based on the sequential Householder transformation. (d) 1D structure satisfying the Hamiltonian $H_{i19}$ in (c). The colors of the connections and elements represent the value of coupling $\kappa$ and the modification of local resonances $\Delta\rho$, respectively. The distributions of $\Delta\rho$ and $\kappa$ are shown in (e). The coupling coefficient in (a,b) is $\kappa_0 = 0.01 \cdot \rho_0$, where the local resonance is normalized as $\rho_0 = 1$.

Because of the difficulty in the removal of crosstalks [28] inside densely packed wave structures, it will be ideal if the transformed tridiagonal Hamiltonian $H_i$ can be achieved with a single, connected structure, when converting a 2D/3D structure with a 1D structure isospectrally. The origin of the disconnection in the 1D design can be elucidated by investigating the graph networks of different degrees of disorder. Figure 2 shows three rewired graphs with different values of the rewiring probability $p$ (Fig. 2a,d,g), their level statics (Fig. 2b,e,h), and transformed 1D designs (Fig. 2c,f,i). For the regular graph of $p = 0$ (Fig. 2a), multilevel edges allow non-unique paths towards effective interactions over the whole network, which derive energy level degeneracy $d$ (Fig. 2b), contrary to the structures only with nearest-neighbor couplings [10]. This emergence of degeneracy originates from the presence of the edge-induced symmetry in the graph (i.e., $d = 5$ for 6 degenerate states, when the degree $k = 6$). Because the finite 1D structure from the tridiagonal $H_i$ lacks enough symmetry for a large degeneracy of the regular network ($d = 5$ in Fig. 2b), a disconnection in the 1D structure is inevitable (black arrows in Fig. 2c), which prohibits the existence of a single and isospectral 1D design for the regular graph. We then impose the disorder upon the regular network [26] to break the structural symmetry from the emergence of shortcut edges and thus lift the level degeneracy dramatically, even for the small value of rewiring probability ($p = 0.05$ in Fig. 2d-f). Most of the networks in the regime between weak (Fig. 2d-f) and completely random (Fig. 2g-i) disorders thus possess their isospectral pair of the 1D structure (Figs 2f and 2i) which is fully connected. We therefore emphasize that the level statistics of hypothetical or physically-not-allowed 'disordered' systems (e.g. 2D/3D systems with significant long-range coupling) can be fully achieved in a 'single' realistic 1D structure of nearest-neighbor evanescent couplings only, through the Householder Hamiltonian transformation $H_i = MHM^T$. Because the regime of strong disorder not only confirms the connected 1D structure more thoroughly but also shows the broadband nature of disorder more evidently (Fig. 2h vs 2e, for the disorder-induced removal of bandgaps [8]), the broadband device based on disordered structures [1,10-13,29] is a suitable application of our network-inspired design methodology.

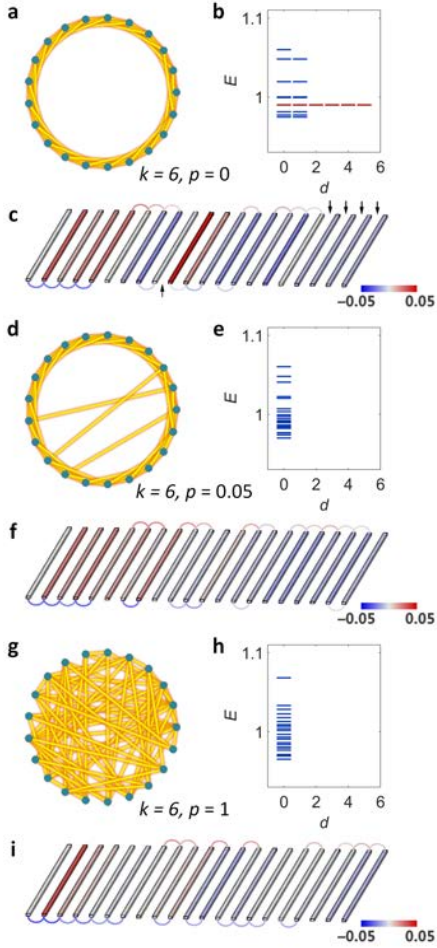

Fig. 2. Lifted degeneracy by disorder in optical systems. (a–c) Regular, (d–f) $p = 0.05$, and (g–i) $p = 1$ rewired networks with their (a,d,g) graphs, (b,e,h) level statistics as a function of the degeneracy $d$, and (c,f,i) transformed 1D structures. Red lines in (b) denote degenerate states. $k = 6$, and the other parameters are same as those in Fig. 1.

A higher-dimensional (2D,3D) structure with a disordered coupling network is the representative example of a high-degree random-walk network. We apply our similarity-based Hamiltonian transformation to a real-space design (Fig. 3): the transformation of a 3D structure that includes all orders of couplings (with higher-order coupling between far-off elements, [19]) into a 1D structure that uses only nearest-neighbor coupling (see Section S1 of Supplement 1 for the effect of higher-order couplings in a 3D real structure). Figure 3a shows the finite-size optical cubic lattice, which possesses periodicity along all Cartesian axes. The disordered structure (Fig. 3b) for the lifted degeneracy is obtained from the 3D random-walk displacement of each atom.

Owing to the spatial variation of the coupling $\kappa$ [17], the corresponding graph for each structure should have $N(N-1)/2$ 'weighted' edges (Fig. 3c,d), which are represented by the varying thicknesses of the lines. The periodicity in the $x$-, $y$-, and $z$-axes in the cubic lattice leads to three symmetric edges with identical coupling strengths (yellow lines in Fig. 3a), which is the origin of the degenerate states. The transformed 1D structure of the cubic lattice is thus divided into three parts ($\kappa = 0$ in Fig. 3e), while the random-walk deformed structure with a randomly weighted graph (Fig. 3d) derives the fully connected isospectral 1D structure (Fig. 3f).

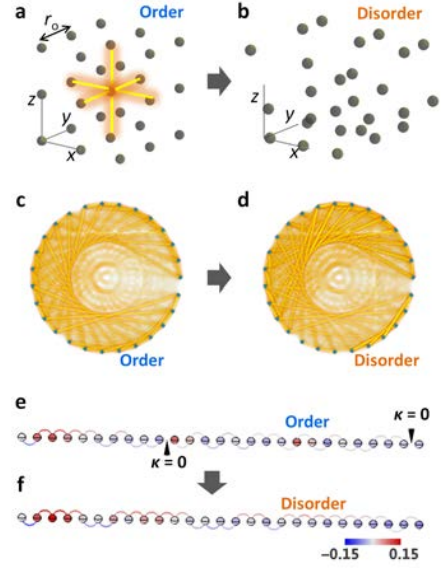

Fig. 3. Network-inspired design in real space. (a) A cubic lattice and (b) its randomly deformed structure. The yellow lines in (a) denote degenerate interactions along the $x$-, $y$-, and $z$-axes, the origin of the level degeneracy. For optical 'atoms', following the exponential relation [17] between the coupling $\kappa$ and the distance between elements $r$, we use the relation of $\kappa = 0.236 \cdot exp(-0.451 \cdot r)$. (c,d) Graph representations and (e,f) transformed 1D structures are shown for the 3D ordered (c,e) and disordered (d,f) structures in (a,b). The original periodicity is $r_o = 5$, the local resonance in 3D structures is $\rho_o = 1$, and the location of each element is deformed by $\Delta_{x,y,z} = 2 \cdot unif(-1,1)$ where $unif(-1,1)$ denotes the uniform distribution between -1 and 1.

Figure 4 shows spectral responses of the network-inspired structures. In line with the results in Fig. 2, the lifted degeneracy (Fig. 4a,b) is achieved from the random-walk deformation. The level repulsion in the eigenspectra for both lattice and deformed structures can be intensified by using weakly-confined local modes, which is desired for broadband applications. To examine the broadband property of the real-space disordered structure, we calculate the absorption spectra with damping elements which have complex self-energy $Im\{\rho_o\} > 0$. The intrinsic $Q$-factor of $Q = 0.5 \cdot Re\{\rho_o\} / Im\{\rho_o\} = 500$ is assumed for each element, to calculate the absorption spectra $A(\rho)$ of the 3D structure and its isospectral 1D partner (see Section S2 of Supplement 1 for the calculation of $A(\rho)$). Figure 4c shows that disordered structures in both 1D and 3D have broader absorption spectra, whereas the transformed 1D structure (symbols) perfectly replicates the original spectrum of each 3D structure (lines). We emphasize that a certain 2D/3D disordered structure with lifted degeneracy [30] can always be simplified to a single 1D structure, which has identical level statistics. As the supersymmetric technique firstly enables the global phase matching for optical multimode structures [20,21,23], it is noted that the inter-dimensional isospectrality provides the global phase matching between multimodes of 2D/3D and 1D optical structures for the first time.

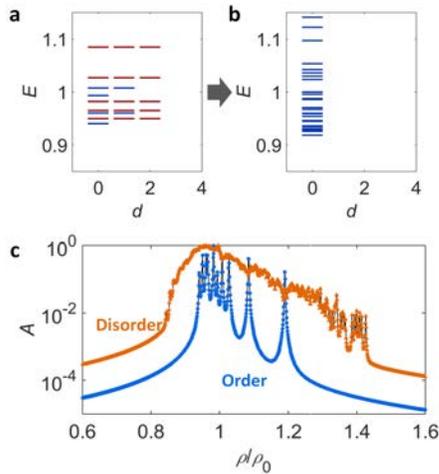

Fig. 4. Spectral responses of network-inspired materials. (a,b) The obtained level statistics for (a) ordered and (b) disordered structures in Fig. 3 are shown as a function of the degeneracy *d*. (c) Normalized absorption spectra of 3D (lines) and 1D (symbols) structures for ordered (blue) and disordered (orange) cases. All spectra are the averages of the statistical ensemble of 200 samples. Each element has the intrinsic quality factor $Q$ = 500.

In summary, in terms of the design technique for optical isospectral structures [8,20-24], we have opened a new direction by exploiting the Hamiltonian similarity in high-degree random-walk networks. Based on previous achievements in the tunable coupling regarding its magnitude [28] or sign [31] in dielectric waveguides or photonic crystal structures, our approach linking wave physics and graph theory can provide a new type of level statistics that originates from graph networks with connected far-off vertices. The fragile degeneracy guarantees the presence of a connected 1D structure that has an identical spectral response to that of a 3D disordered structure. Our results thus extend the field of correlated disorder [1,2,8,10,14,32] into network-like level statistics.

Our approach utilizing the Hamiltonian transformation allows the finding of 1D isospectral structures from random networks deterministically, in contrast to numerical searching techniques, *e.g.* genetic algorithm. We also note that the impact of higher-order coupling is critical for the analysis of densely-packed and large-scale devices [19,33], and the engineering of spectral singularity [14] and slow light dispersion [34]. Our deterministic 1D realization thus provides simple and controllable achievements of higher-order coupling effects. Furthermore, from the inter-dimensional global phase matching, efficient wave transfers between different dimensions can be achieved (e.g. 2D-to-1D coupler in Section S3 of Supplement 1).

**Funding.** National Research Foundation of Korea (NRF) (NRF-2014M3A6B3063708), (K20815000003), (2016R1A6A3A04009723).

**Acknowledgment**. The authors thank anonymous reviewers for their valuable comments and insights.

See Supplement 1 for supporting content.

# Inter-dimensional optical isospectrality inspired by graph networks: supplementary material

Sunkyu Yu,[1] Xianji Piao,[1] Jiho Hong,[1] and Namkyoo Park[1,*]

[1]*Photonic Systems Laboratory, Department of Electrical and Computer Engineering, Seoul National University, Seoul 08826, Korea*
**Corresponding author:* nkpark@snu.ac.kr*

**This document provides supplementary information to "Inter-dimensional optical isospectrality inspired by graph networks," http://dx.doi.org/XX/XXXX/optica.X.XXXXX. Additional details are given about the effects of higher-order couplings, the calculation of absorption spectra, and the multimode coupling between 1D and 2D structures.**

### S1. Effects of higher-order couplings

Here we analyze the effect of higher-order couplings [1] in 3D real structures. Figure S1 shows the graph representations (Fig. S1a-c) and level statistics (Fig. S1d-f) of the ordered structure in Fig. 3a in the main manuscript, for the inclusion of nearest-neighbor (NN) coupling (Fig. S1a,d), NN and next-nearest-neighbor (NNN) couplings (Fig. S1b,e), and all orders of couplings (Fig. S1c,f). Despite their smaller values, higher-order couplings have a strong impact on the level statistics (Fig. S1d-f). Higher-order couplings also reduce the level degeneracy due to the broken symmetry from different values for each order of coupling, in contrast to the case of hypothetical graph systems with identical multilevel edges which derive the level degeneracy (Fig. 2a in the main manuscript).

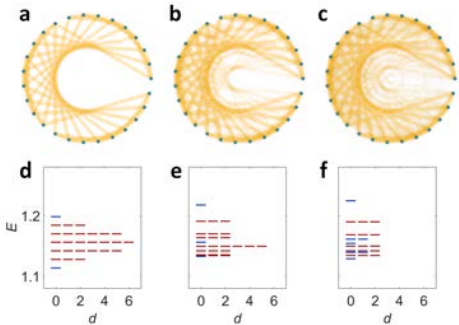

Fig. S1. Effects of higher-order coupling on the level statistics of the ordered structure. (a-c) Graphs and (d-f) level statistics for the inclusion of (a,d) NN coupling, (b,e) NN and NNN couplings, and (c,f) all orders of couplings for the ordered structure in Fig. 3a.

Figure S2 represents the case of the disordered structure in Fig. 3b in the main manuscript. Figure S2a,d includes only the couplings smaller than that between elements with the distance $r_o$, while Fig. S2b,e includes the couplings smaller than that between elements with the distance $2^{1/2} \cdot r_o$ (all orders of couplings for Fig. S2c,f). As seen, the change of the level statistics according to the considered orders of couplings is apparent.

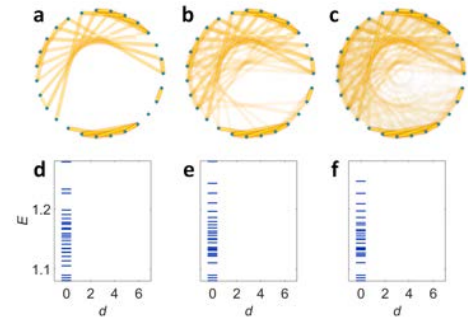

Fig. S2. Effects of higher-order coupling on the level statistics of the disordered structure. (a-c) Graphs and (d-f) level statistics for the inclusion of (a,d) NN coupling, (b,e) NN and NNN couplings, and (c,f) all orders of couplings for the disordered structure in Fig. 3b.

### S2. Calculation of absorption spectra

To obtain absorption spectra of the 3D structure and its isospectral 1D partner, we assume lossy elements ($Im\{\rho_o\} > 0$) of the 3D one. Each element is assumed to have the intrinsic $Q$-factor of $Q = 0.5 \cdot Re\{\rho_o\} / Im\{\rho_o\} = 500$. From the eigenvalue equation $H\psi = E\psi$, we can obtain the set of complex eigenvalues $E_j$ ($j = 1, 2, ..., N$). The corresponding absorption spectrum [2] $A(\rho)$ neglecting the interference between eigenmodes then becomes

$$A(\rho) = \sum_{j=1}^{N} \frac{1}{\left[\rho - Re\{E_j\}\right]^2 + \left[Im\{E_j\}\right]^2}.$$

The absorption spectrum of the 1D partner potential can also be obtained from the eigenvalue equation $H_i\psi = E\psi$ where $H_i = MHM^T$. Figure 4c in the main manuscript shows the normalized absorption spectra of the 3D structure (lines) and its isospectral 1D partner (symbols).

### S3. Inter-dimensional multimode coupling

The isospectrality in multimode structures provide the global phase matching condition [3,4]: the simultaneous matching of eigenvalues, allowing the power transfer for all of multimodes. As an application example, here we demonstrate the multimode coupling between different dimensions (2D-to-1D) for the first time, by utilizing the global phase matching condition from the inter-dimensional isospectrality. Figure S3a shows the proposed structure. From the original 2D structure composed of identical optical atoms (blue dashed circle), we achieve the 1D isospectral partner structure (red dashed ellipse), by following the transformation $H_i = MHM^T$ in the main manuscript (Fig. S3b for the self-energy and coupling strength distribution of the 1D structure; Fig. S3c for the identical level statistics of both structures). We note that 2D and 1D structures are coupled through a single atom of each structure, with the coupling coefficient $\kappa_d$. To preserve the global phase matching condition, the strength of $\kappa_d$ is smaller than internal couplings in 2D and 1D structures as $\kappa_d < \kappa_{ij}$.

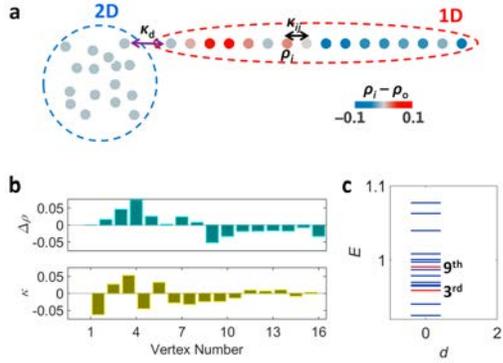

Fig. S3. The design of the inter-dimensional multimode coupling. (a) A schematic for 2D and 1D structures. For the optical atoms in the 2D structure, we use the same parameters as those in Fig. 3 in the main manuscript: $\kappa = 0.236 \cdot exp(-0.451 \cdot r)$, $r_o = 5$, $\rho_o = 1$, and $\Delta_{x,y,z} = 2 \cdot unif(-1,1)$. The inter-dimensional coupling $\kappa_d = 0.01$. The colors of the elements represent the value of the modification of local resonances $\Delta\rho = \rho - \rho_o$. (b) The distributions of $\Delta\rho$ and $\kappa$ in the 1D structure. (c) The level statistics of both structures. Red eigenstates for Fig. S4.

Figure S4 shows the coupling between 2D and 1D structures. As an example, the 3rd and 9th eigenstates are excited in the 2D structure. Due to the matching of each eigenvalue, full couplings between different dimensions are successfully achieved for both eigenstates (3rd state: Fig. S4a, 9th state: Fig. S4b). Figure S4c-S4e (S4f-S4h) presents the field distribution for the excitation of the 3rd (9th) eigenstate in the 2D structure. We also note that because the effective coupling between structures is dependent on each modal profile, the beat length is different for each mode.

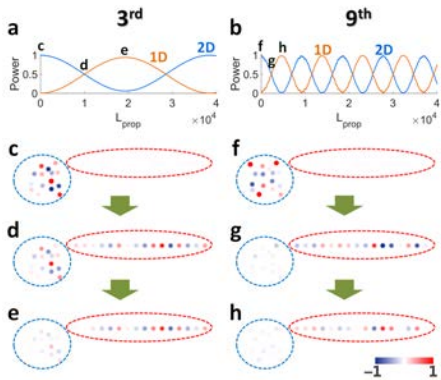

Fig. S4. Inter-dimensional coupling between 2D and 1D structures. Waveguide optical atoms are assumed, for the implementation of the directional coupling during the propagation. (a,b) Total power flow in each structure for the propagation length $L_{prop}$. (c-h) Field distributions in both structures at the propagation lengths depicted in (a,b). (a,c-e) 3rd state, (b,f-h) 9th state.